\documentclass[traditabstract,rnote]{aa}  

\usepackage{graphicx}
\usepackage{txfonts}
\usepackage{epsfig}
\usepackage{natbib}

\begin{document}

\title{The host galaxy of 3C 279}

\author{
K. Nilsson\inst{1}\and
T. Pursimo\inst{2}\and
C. Villforth\inst{1}\fnmsep\inst{2}\and
E. Lindfors\inst{1}\and
L. O. Takalo\inst{1}
}
   
\institute{Tuorla Observatory, Department of Physics and Astronomy,
University of Turku, V\"ais\"al\"antie 20, FI-21500 Piikki\"o, Finland
\and
Nordic Optical Telescope, Apartado 474, 38700 Santa Cruz de La
Palma, Spain}

\date{Received; accepted}

\abstract { We have obtained a deep i-band image of the blazar 3C 279
while the target was in a low optical state. Due to the faintness of
the optical nucleus we have made the first detection of the host
galaxy. The host galaxy has an apparent I-band magnitude of
18.4$\pm$0.3 and an effective radius of (2.7 $\pm$ 1.1) arcsec.  The
luminosity of the host galaxy $M_R$ = -23.8 is consistent with
the luminosities of other radio-loud quasar host galaxies. Using the
empirical correlation between bulge luminosity and central black hole
mass $M_{bh}$ we estimate $\log(M_{bh}/M_{\odot}) = 8.9\pm0.5$,
broadly consistent with values obtained by photoionization methods.}


\maketitle

\section{Introduction}

The host galaxies of active galactic nuclei (AGN) provide important
information of the environment and evolution of their central black
holes.  The close connection between the central black holes and their
host galaxies is clearly manifested by the correlations between the
central black hole mass $M_{bh}$ and the host galaxy parameters,
such as the central velocity dispersion $\sigma$
\citep{2000ApJ...539L...9F,2000ApJ...539L..13G,2001ApJ...547..140M},
bulge mass $M_{bulge}$ \citep{1998AJ....115.2285M} and galaxy light
concentration $C_{re}(1/3)$ \citep{2001ApJ...563L..11G}. It is usually
much easier to determine the host galaxy parameters than to apply
kinematic methods or reverberation mapping
\citep[e.g.][]{2000ApJ...533..631K} making the host galaxy method an
appealing alternative to estimate black hole masses in galaxies.  Host
galaxies can also be used to study the unified schemes of AGN since
their properties do not depend on viewing angle
\citep{1995PASP..107..803U}.

\object{3C 279} (z = 0.538) is an AGN which belongs to the class of
blazars, whose most notable features are strong continuum radiation
and high variability over the whole electromagnetic spectrum and high
optical and radio polarization. The blazar class comprises of flat
spectrum radio quasars (FSRQs) and BL Lacertae objects (BL Lacs) whose
main difference in the optical lies in the strength of their broad
emission lines: the latter group has very weak or nonexistent emission
lines whereas in the former they are more pronounced. \object{3C 279}
is classified as a FSRQ, although both the emission line strengths and
$\gamma$-ray properties place it close to the FSRQ-BL Lac border
\citep{1998ApJ...492..173K,2009MNRAS.396L.105G}.

Despite of \object{3C 279} being subject to intensive study over the
last three decades, determining its host galaxy parameters has
received little attention.  The only published host galaxy study was
made by \cite{1998A&A...332..503K} who observed \object{3C 279} in the
H-band, but could only derive an upper limit of H $>$ 13.7 for the
host galaxy. One of the main factors affecting the host galaxy
detection efficiency is the nucleus to host ratio: when the nucleus is
very bright compared to the host galaxy its is difficult to detect the
host galaxy, let alone reliably determine its main parameters
(magnitude and effective radius).  \object{3C 279} is known to be
highly variable in the optical bands with historical variations over 6
magnitudes in the B-band \citep{1990AJ....100.1452W}.  The rapid
variability of \object{3C 279} makes the host galaxy detection even
more challenging as is it difficult to observe the target in a low
state.

We are currently monitoring over 40 blazars in the R-band in the
Tuorla Observatory blazar monitoring
program\footnote{http://users.utu.fi/$\sim$kani/1m/index.html}.  In
the end of June 2008 we observed \object{3C 279} to go into a fairly
deep optical minimum (R $\sim$ 16.8) and initiated prompt i-band
imaging at the Nordic Optical Telescope (NOT) to detect its host
galaxy. The results of this imaging are presented in this research
note.

Throughout this paper we use the cosmology $H_0 = 70$ km s$^{-1}$
Mpc$^{-1}$, $\Omega_{M}$ = 0.3 and $\Omega_{\Lambda}$ = 0.7.

\section{Observations and data reduction}

\begin{figure}
\begin{center}
\epsfig{file=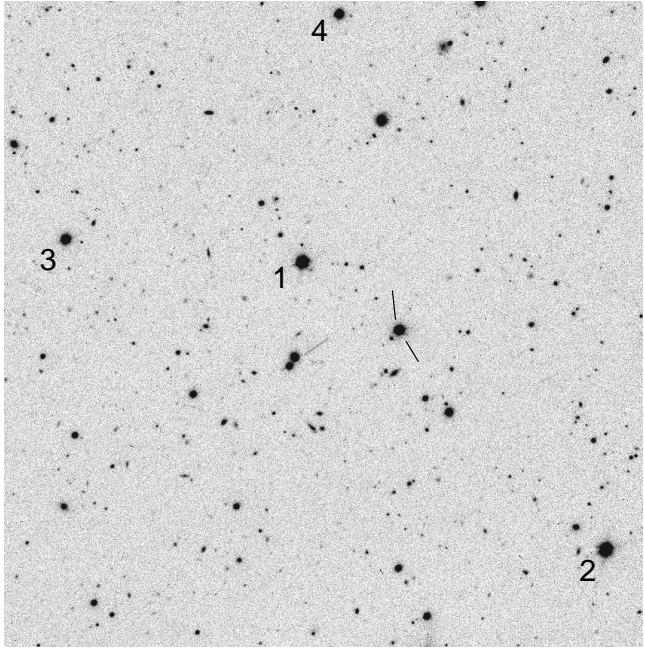,width=8.5cm}
\end{center}
\caption{\label{kentta} The summed i-band image of the field around
\object{3C 279}. The field size is 5.3'$\times$5.4', North is up and
East is to the left. Stars 1-4 are discussed in the text.}
\end{figure}

The observations were made in June 23-24, 2008 with the Nordic Optical
Telescope (NOT) using the ALFOSC instrument. This instrument has a
pixel scale of 0\farcs189/pixel and a gain of 0.726 $e^-$/ADU and a
readout noise of 3.2 electrons. The field of view of the instrument is
6.5' $\times$ 6.5'. Altogether 24 images of the \object{3C 279} field
were obtained through an i-band filter with almost an uniform
transmission between 725 and 875 nm. Individual exposure times were
kept low enough (100-200s) to leave \object{3C 279} and a nearby star
(star 1 in Fig.  \ref{kentta}) unsaturated. The position of
\object{3C 279} on the CCD was changed between exposures to be able
to make a fringe correction image.

The individual images were first bias-subtracted and then flat-fielded
with twilight flats in the usual way using IRAF\footnote{ IRAF is
distributed by the National Optical Astronomy Observatories, which are
operated by the Association of Universities for Research in Astronomy,
Inc., under cooperative agreement with the National Science
Foundation.}. The fringe pattern visible in the images was corrected
by subtracting a fringe correction image produced by median combining
all 24 images and applying simultaneously a sigma clipping cut to the
pixel intensity values. After correction for the fringe pattern, the
individual frames were registered and summed.  The summed image shown
in Fig. \ref{kentta} has a FWHM of 0.79 arcsec and a total exposure
time of 3300 s.

Calibration of the field was obtained via star 1 in Fig. \ref{kentta},
for which \cite{1998PASP..110.1164S} give I = 15.00 $\pm$ 0.04. Since
the i-band filter used in our observations closely matches the Cousins
I-band filter, we expect only small color effects between our i-band
magnitudes and standard I-band magnitudes. We conservatively estimate
the error of calibration to be 0.1 mag. 

\section{Analysis and results}

We looked for the host galaxy of \object{3C 279} by fitting
two-dimensional surface brightness models to the observed i-band
image. We used three different models: an unresolved nucleus only,
nucleus + a de Vaucouleurs profile host galaxy (S\'ersic $\beta$ =
0.25) and nucleus + a disk host galaxy ($\beta$ = 1.0). Details of
this fitting process can be found in \cite{1999PASP..111.1223N}. In
short, the model has three adjustable parameters, the magnitude of the
nucleus $I_{\rm core}$, the magnitude of the host galaxy $I_{\rm
  host}$ and the effective radius of the host galaxy $r_{\rm eff}$.
The unresolved nucleus is assumed to be centered on the host galaxy.
The three parameters are adjusted via an iterative
Levenberg-Marquardt loop until the minimum chi squared between the
model and the observed surface brightness distribution is found. The
fit was extended to an outer radius of 9\farcs5, except that the
pixels affected by a galaxy 5\farcs9 SE of \object{3C 279} were
masked out from the fit.

Prior to computing the chi squared the model was convolved with the
PSF. We used star 1 in Fig. \ref{kentta} for the PSF due to its high
signal to noise and proximity to 3C 279.  As in
\cite{2008A&A...487L..29N}, we studied the variability of the PSF
across the FOV by extracting the surface brightness profiles of stars
1-4 in Fig. \ref{kentta} and deriving the rms scatter between the
profiles. The PSF variability was then parameterized by a parabolic
expression
\begin{equation}
\sigma_{\rm PSF}(r) = 0.035 + 2 \times 10^{-5} \cdot r^2\ ,
\end{equation}
where $\sigma_{\rm PSF}(r)$ is the uncertainty in the PSF relative to
the intensity at radius $r$ (pixels) from the PSF center. We then
computed the expected variance $\sigma^2$ in a pixel with intensity
$I$ (ADU) and distance $r$ from the center of \object{3C 279} from
the expression
\begin{equation}
\label{nkaava}
\sigma^2 = \frac{G*I+R^2}{G^2} + [\sigma_{\rm PSF}(r) * I]^2\ ,
\end{equation}
where $G$ is the effective gain and $R$ is the effective readout noise,
and compute the $\chi^2$ from
\begin{equation}
\chi^2 = \sum_i \frac{(I_i - M_i)^2}{\sigma_i^2}\ ,
\end{equation}
where $M$ is the model intensity and the summation is over all
unmasked pixels within the fitting radius. By including the PSF
variability into the computation of $\chi^2$ we can ensure that the
results are not dominated by PSF errors, which are most pronounced
close to the center of the object.

\begin{figure}
\begin{center}
\epsfig{file=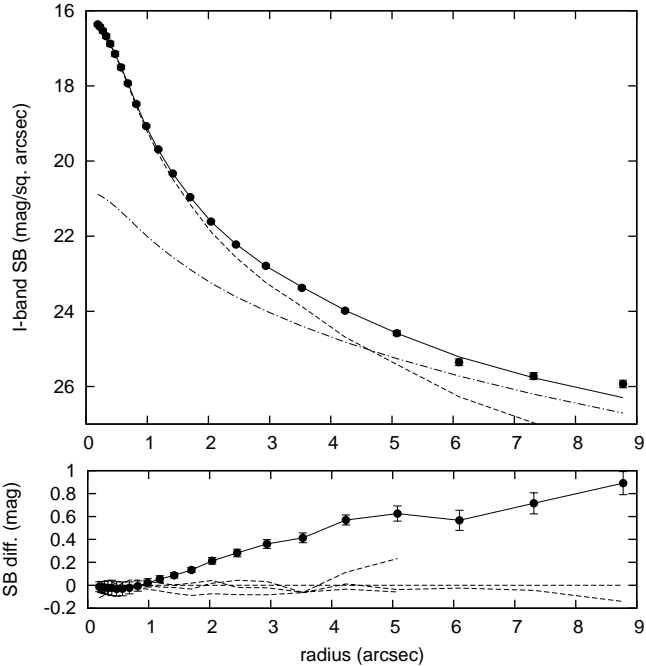,width=9cm}
\end{center}
\caption{\label{profiilit} {\em Upper panel}: The surface brightness
profile of \object{3C 279} (filled symbols) together with the model
(solid line), the nucleus (dashed line) and the host galaxy
(dot-dashed line). {\em Lower panel}: The surface brightness profiles
of \object{3C 279} (solid line and filled symbols) and stars 1-4 in
Fig. \ref{kentta} relative to star 1. Each profile is traced out to
a radius where the surface brightness can be determined to better
that 10\% accuracy. Note that star 1 is represented by a horizontal
line at SB = 0.0 in the lower panel.}
\end{figure}

The results of this model fitting are summarized in Table
\ref{tulokset} and Fig. \ref{profiilit}. We see a clear excess in 3C
279 over the PSF and this excess is clearly above any PSF variability
across the FOV (see the lower panel of Fig. \ref{profiilit}). The
models with a host galaxy give a much better fit than a model with
pure nucleus with $\chi^2$ indicating a near-perfect fit. The model
with a de Vaucouleurs host galaxy gives a slightly better fit that the
model with a disk galaxy but the difference is not significant
according to our error simulations (see below). However, since the de
Vaucouleurs profile gave formally a better fit and no blazar host
galaxy has ever been associated with a disk galaxy, we concentrate in
the following to the results with the de Vaucouleurs profile.

\begin{table}
\caption{\label{tulokset} The results of the model fitting.}
\centering
\begin{tabular}{lcccc}
\hline
\hline
Model & I$_{\rm core}$ & I$_{\rm host}$ & $r_{\rm eff}$ & $\chi^2$/dof\\
\hline
Nucleus         & 15.99 &       &          & 1.38\\
Nucleus + De V. & 16.13 & 18.43 & 2\farcs7 & 1.03\\ 
Nucleus + disk  & 16.08 & 19.04 & 2\farcs8 & 1.05\\
\hline
\end{tabular}
\end{table}

We have studied the sensitivity of the results to random noise, PSF
variability and errors in the assumed profile of the host galaxy by
performing 50 model fits to simulated images of \object{3C 279}.  In
these simulations $I_{\rm core}$, $I_{\rm host}$ and $r_{\rm eff}$
were held constant at the values shown in Table \ref{tulokset}, but
the contribution from the different noise sources changed from one
simulation to another.

The random noise in these simulations is assumed to raise from readout
and photon noise according to the first term on the right hand side of
Eq. \ref{nkaava}. To model the PSF variability we created two PSF
models in each simulation, slightly differing from each other. The
simulated model was convolved with the first PSF and the model fits
were made with the second PSF. Both PSFs consisted of an elliptical
Moffat profile with $\beta$ = 2.5.  The ellipticity of the first PSF
was randomly drawn from an uniform distribution with a minimum of 0.0
and a maximum of 0.12 and the position angle from an uniform
distribution between 0 and 180 degrees.  For the second PSF the
ellipticity was drawn from a Gaussian distribution with a mean equal
to the ellipticity of the first PSF and $\sigma = 0.05$.  The position
angle of the second PSF was similarly drawn from a Gaussian
distribution with a mean equal to the first PSF and $\sigma = 12.5$
degrees. The limits and standard deviations above were chosen to mimic
the range of PSF shapes seen in our i-band images and to roughly
reproduce the residuals seen in the model subtracted image of
\object{3C 279}.

We also included in the simulations the possibility that the host
galaxy does not exactly follow the de Vaucouleurs profile with the
slope of the surface brightness profile $\beta$ equal to 0.25, as
often is observed \citep[e.g.][]{1999PASP..111.1223N}. Therefore, when
creating the host galaxy model the profile slope $\beta$ of the host
galaxy was drawn from a Gaussian distribution with mean = 0.25 and
$\sigma = 0.06$. When performing the fits to the simulated images the
$\beta$ parameter was held constant at 0.25, however.

After completing the simulations we computed the standard deviations
of $I_{\rm core}$, $I_{\rm host}$ and $r_{\rm eff}$ to be 0.02 mag,
0.18 mag and 1.1 arcsec, respectively. Taking into account the error
in the calibration we end up with $I_{\rm host}$ = 18.4 $\pm$ 0.3 and
$r_{\rm eff}$ = 2.7 $\pm$ 1.1 arcsec for the host galaxy. The $\chi^2$
in the simulated fits had an average of 1.11 and standard deviation
$\sigma = 0.08$. Given that the scatter in $\chi^2$ in the simulations
is significantly larger than the difference in $\chi^2$ between
the de Vaucouleurs and disk models, we conclude that we cannot claim
either of the two models to be significantly better.

\section{Discussion}

Since most host galaxy parameters and correlations have been discussed
in the R-band we first convert the host galaxy magnitude from I-band
to R-band. For the conversion we used $A_I$ = 0.06 for the galactic
extinction \citep{1998ApJ...500..525S}, $K_I + e(I)$ = 0.4 for the
K-correction and evolutionary correction, based on
\cite{1998MNRAS.300..872M} and \cite{2000ApJ...541..126M} and assuming
galaxy formation at z = 2 and finally R - I = 0.7 for an elliptical
galaxy \citep{1995PASP..107..945F}.  Using these values we derive R =
18.7 for the host galaxy of \object{3C 279}, which translates into
$M_R$ = -23.8 in the adopted cosmology.  This value is in the bright
end of the values found for the host galaxies of radio-loud quasars
\citep[-23.2 $\pm$ 0.3,][transferred to the cosmology used
here]{2003MNRAS.340.1095D} and BL Lacertae objects \citep[-22.9 $\pm$
0.5,][]{2005ApJ...635..173S}. The effective radius is (17 $\pm$ 7)
kpc, which is fairly large compared to typical effective radii ($\sim$
10 kpc) found for radio-loud quasars and BL Lacs, although the margin
of error is also large.

Next we estimate the central black hole mass of \object{3C 279}
using the relationship between bulge luminosity and central black hole
mass derived in \cite{2009ApJ...694L.166B} using the same cosmology as
here: $ \log(M_{bh}/10^8 M_{\odot}) = -0.02 + 0.8 \log(L_{\rm bulge} /
10^{10} L_{\odot}) $, where $L_{\rm bulge}$ is measured in the V-band
This relationship was derived using 26 Seyfert 1 galaxies and quasars
at z $<$ 0.3. Although the redshift of \object{3C 279} is outside this
range, the host galaxy luminosity of \object{3C 279} is within the
range of V-band bulge luminosities probed by this study $9 <
\log(L_{\rm bulge}/L_{\odot}) < 11.5$.

As blazar host galaxies are expected to be bulge-dominated systems we
use the derived R-band magnitude directly for computing the bulge
luminosity. We first transferred the R-band luminosity of the host
galaxy to the V-band using V - R = 0.6 \citep{1995PASP..107..945F} and
used the above relation to derive $\log(M_{bh}/M_{\odot})$ = 8.9 for
\object{3C 279}.  The scatter in the $L_{\rm bulge}-M_{bh}$ relation
is $\sim$ 0.4 dex. The uncertainty arising from the host galaxy
magnitude error is smaller ($<$ 0.2 dex), so our estimate of the black
hole mass has a total error of $\sim$ 0.5 dex. Like the host galaxy
luminosity, the derived black hole mass is in the upper end of the
black hole mass distribution of FSRQs, but still consistent with the
total observed range \citep[see e.g.][]{2008MNRAS.387.1669G}.

\begin{table}
\caption{\label{massat} Previous black hole mass determinations for
3C 279.}
\begin{center}
\begin{tabular}{llc}
\hline
\hline
$\log(M_{bh}/M_{\odot})$ & method & ref.\\
\hline
8.912 & FWHM (H$_{\beta}$) - $L_{opt}$ & 1\\
9.099 & FWHM (H$_{\beta}$) - $L_{opt}$ & 2\\
8.43  & FWHM (H$_{\beta}$) - $L_{opt}$ & 3\\
8.6   & $\gamma$-ray luminosity        & 4\\
8.48  & FWHM (H$_{\beta}$) - $L_{opt}$ & 5\\
9.79  & FWHM (H$_{\beta}$) - $L_{opt}$ & 6\\
8.28  & FWHM (H$_{\beta}$) - $L_{opt}$ & 7\\
\hline
\end{tabular}
\end{center}
References: 
(1) \cite{2001MNRAS.327.1111G}, 
(2) \cite{2002MNRAS.331..111C},
(3) \cite{2002ApJ...579..530W},
(4) \cite{2003MNRAS.340..632L},
(5) \cite{2004ApJ...615L...9W}, 
(6) \cite{2004MNRAS.347..607B},
(7) \cite{2006ApJ...637..669L}
\end{table}

We have looked in the literature for other black hole mass
determinations for \object{3C 279}. The results of this search are
summarized in Table \ref{massat}. It is beyond the scope of this paper
to discuss in detail the various methods used to derive the values in
Table \ref{massat} and their relative merits. Most of them have been
determined with the empirical FWHM (H$_{\beta}$) - $L_{opt}$
correlation \citep[e.g.][]{1999ApJ...526..579W} with different
assumptions of BLR geometry. There is a broad range of values found by
different authors, but most of them are in the range $8.5 <
\log(M_{bh}/M_{\odot}) <9.0$, broadly consistent with the value found
here.  The two last values in Table \ref{massat} deviate most from the
general consensus and at least in \cite{2004MNRAS.347..607B} (ref. 6 in
table \ref{massat}) the high $M_{bh}$ estimate can be traced to a much
broader reported H$_{\beta}$ line width than the other authors.

\section{Summary}

The results of this paper can summarized as follows:

(1) We have obtained a deep i-band image of 3C 279 which
has enabled us to reveal the host galaxy for the first time.

(2) The luminosity of the host galaxy ($M_R$ = -23.8) is in the bright
end of luminosities of other radio-loud quasar host galaxies.  The
derived effective radius (17 $\pm$ 7 kpc) is quite large compared to
other radio-loud quasars, but the effective radius is not very well
constrained.

(3) Using the empirical correlation between bulge luminosity and
central black hole mass $M_{bh}$ we estimate $\log(M_{bh}/M_{\odot}) =
8.9\pm0.5$, broadly consistent with values obtained by photoionization
methods.

\begin{acknowledgements}

The authors thank Talvikki Hovatta for useful discussions during the
preparation of this paper.  These data are based on observations made
with the Nordic Optical Telescope, operated on the island of La Palma
jointly by Denmark, Finland, Iceland, Norway, and Sweden, in the
Spanish Observatorio del Roque de los Muchachos of the Instituto de
Astrofisica de Canarias.  The data presented here have been taken
using ALFOSC, which is owned by the Instituto de Astrofisica de
Andalucia (IAA) and operated at the Nordic Optical Telescope under
agreement between IAA and the NBIfAFG of the Astronomical Observatory
of Copenhagen.This research has made use of the NASA/IPAC
Extragalactic Database (NED) which is operated by the Jet Propulsion
Laboratory, California Institute of Technology, under contract with
the National Aeronautics and Space Administration.

\end{acknowledgements}

\bibliographystyle{aa}

\bibliography{12820.bib}

\begin{thebibliography}{30}
\expandafter\ifx\csname natexlab\endcsname\relax\def\natexlab#1{#1}\fi

\bibitem[{{Bentz} {et~al.}(2009){Bentz}, {Peterson}, {Pogge}, \&
  {Vestergaard}}]{2009ApJ...694L.166B}
{Bentz}, M.~C., {Peterson}, B.~M., {Pogge}, R.~W., \& {Vestergaard}, M. 2009,
  \apjl, 694, L166

\bibitem[{{Bian} \& {Zhao}(2004)}]{2004MNRAS.347..607B}
{Bian}, W. \& {Zhao}, Y. 2004, \mnras, 347, 607

\bibitem[{{Cao} \& {Jiang}(2002)}]{2002MNRAS.331..111C}
{Cao}, X. \& {Jiang}, D.~R. 2002, \mnras, 331, 111

\bibitem[{{Dunlop} {et~al.}(2003){Dunlop}, {McLure}, {Kukula}, {Baum}, {O'Dea},
  \& {Hughes}}]{2003MNRAS.340.1095D}
{Dunlop}, J.~S., {McLure}, R.~J., {Kukula}, M.~J., {et~al.} 2003, \mnras, 340,
  1095

\bibitem[{{Ferrarese} \& {Merritt}(2000)}]{2000ApJ...539L...9F}
{Ferrarese}, L. \& {Merritt}, D. 2000, \apjl, 539, L9

\bibitem[{{Fukugita} {et~al.}(1995){Fukugita}, {Shimasaku}, \&
  {Ichikawa}}]{1995PASP..107..945F}
{Fukugita}, M., {Shimasaku}, K., \& {Ichikawa}, T. 1995, \pasp, 107, 945

\bibitem[{{Gebhardt} {et~al.}(2000){Gebhardt}, {Bender}, {Bower}, {Dressler},
  {Faber}, {Filippenko}, {Green}, {Grillmair}, {Ho}, {Kormendy}, {Lauer},
  {Magorrian}, {Pinkney}, {Richstone}, \& {Tremaine}}]{2000ApJ...539L..13G}
{Gebhardt}, K., {Bender}, R., {Bower}, G., {et~al.} 2000, \apjl, 539, L13

\bibitem[{{Ghisellini} {et~al.}(2009){Ghisellini}, {Maraschi}, \&
  {Tavecchio}}]{2009MNRAS.396L.105G}
{Ghisellini}, G., {Maraschi}, L., \& {Tavecchio}, F. 2009, \mnras, 396, L105

\bibitem[{{Ghisellini} \& {Tavecchio}(2008)}]{2008MNRAS.387.1669G}
{Ghisellini}, G. \& {Tavecchio}, F. 2008, \mnras, 387, 1669

\bibitem[{{Graham} {et~al.}(2001){Graham}, {Erwin}, {Caon}, \&
  {Trujillo}}]{2001ApJ...563L..11G}
{Graham}, A.~W., {Erwin}, P., {Caon}, N., \& {Trujillo}, I. 2001, \apjl, 563,
  L11

\bibitem[{{Gu} {et~al.}(2001){Gu}, {Cao}, \& {Jiang}}]{2001MNRAS.327.1111G}
{Gu}, M., {Cao}, X., \& {Jiang}, D.~R. 2001, \mnras, 327, 1111

\bibitem[{{Kaspi} {et~al.}(2000){Kaspi}, {Smith}, {Netzer}, {Maoz}, {Jannuzi},
  \& {Giveon}}]{2000ApJ...533..631K}
{Kaspi}, S., {Smith}, P.~S., {Netzer}, H., {et~al.} 2000, \apj, 533, 631

\bibitem[{{Koratkar} {et~al.}(1998){Koratkar}, {Pian}, {Urry}, \&
  {Pesce}}]{1998ApJ...492..173K}
{Koratkar}, A., {Pian}, E., {Urry}, C.~M., \& {Pesce}, J.~E. 1998, \apj, 492,
  173

\bibitem[{{Kotilainen} {et~al.}(1998){Kotilainen}, {Falomo}, \&
  {Scarpa}}]{1998A&A...332..503K}
{Kotilainen}, J.~K., {Falomo}, R., \& {Scarpa}, R. 1998, \aap, 332, 503

\bibitem[{{Liang} \& {Liu}(2003)}]{2003MNRAS.340..632L}
{Liang}, E.~W. \& {Liu}, H.~T. 2003, \mnras, 340, 632

\bibitem[{{Liu} {et~al.}(2006){Liu}, {Jiang}, \& {Gu}}]{2006ApJ...637..669L}
{Liu}, Y., {Jiang}, D.~R., \& {Gu}, M.~F. 2006, \apj, 637, 669

\bibitem[{{Magorrian} {et~al.}(1998){Magorrian}, {Tremaine}, {Richstone},
  {Bender}, {Bower}, {Dressler}, {Faber}, {Gebhardt}, {Green}, {Grillmair},
  {Kormendy}, \& {Lauer}}]{1998AJ....115.2285M}
{Magorrian}, J., {Tremaine}, S., {Richstone}, D., {et~al.} 1998, \aj, 115, 2285

\bibitem[{{Maraston}(1998)}]{1998MNRAS.300..872M}
{Maraston}, C. 1998, \mnras, 300, 872

\bibitem[{{Maraston} \& {Thomas}(2000)}]{2000ApJ...541..126M}
{Maraston}, C. \& {Thomas}, D. 2000, \apj, 541, 126

\bibitem[{{Merritt} \& {Ferrarese}(2001)}]{2001ApJ...547..140M}
{Merritt}, D. \& {Ferrarese}, L. 2001, \apj, 547, 140

\bibitem[{{Nilsson} {et~al.}(2008){Nilsson}, {Pursimo}, {Sillanp{\"a}{\"a}},
  {Takalo}, \& {Lindfors}}]{2008A&A...487L..29N}
{Nilsson}, K., {Pursimo}, T., {Sillanp{\"a}{\"a}}, A., {Takalo}, L.~O., \&
  {Lindfors}, E. 2008, \aap, 487, L29

\bibitem[{{Nilsson} {et~al.}(1999){Nilsson}, {Pursimo}, {Takalo},
  {Sillanp{\"a}{\"a}}, {Pietil{\"a}}, \& {Heidt}}]{1999PASP..111.1223N}
{Nilsson}, K., {Pursimo}, T., {Takalo}, L.~O., {et~al.} 1999, \pasp, 111, 1223

\bibitem[{{Sbarufatti} {et~al.}(2005){Sbarufatti}, {Treves}, \&
  {Falomo}}]{2005ApJ...635..173S}
{Sbarufatti}, B., {Treves}, A., \& {Falomo}, R. 2005, \apj, 635, 173

\bibitem[{{Schlegel} {et~al.}(1998){Schlegel}, {Finkbeiner}, \&
  {Davis}}]{1998ApJ...500..525S}
{Schlegel}, D.~J., {Finkbeiner}, D.~P., \& {Davis}, M. 1998, \apj, 500, 525

\bibitem[{{Smith} \& {Balonek}(1998)}]{1998PASP..110.1164S}
{Smith}, P.~S. \& {Balonek}, T.~J. 1998, \pasp, 110, 1164

\bibitem[{{Urry} \& {Padovani}(1995)}]{1995PASP..107..803U}
{Urry}, C.~M. \& {Padovani}, P. 1995, \pasp, 107, 803

\bibitem[{{Wandel} {et~al.}(1999){Wandel}, {Peterson}, \&
  {Malkan}}]{1999ApJ...526..579W}
{Wandel}, A., {Peterson}, B.~M., \& {Malkan}, M.~A. 1999, \apj, 526, 579

\bibitem[{{Wang} {et~al.}(2004){Wang}, {Luo}, \& {Ho}}]{2004ApJ...615L...9W}
{Wang}, J.-M., {Luo}, B., \& {Ho}, L.~C. 2004, \apjl, 615, L9

\bibitem[{{Webb} {et~al.}(1990){Webb}, {Carini}, {Clements}, {Fajardo},
  {Gombola}, {Leacock}, {Sadun}, \& {Smith}}]{1990AJ....100.1452W}
{Webb}, J.~R., {Carini}, M.~T., {Clements}, S., {et~al.} 1990, \aj, 100, 1452

\bibitem[{{Woo} \& {Urry}(2002)}]{2002ApJ...579..530W}
{Woo}, J.-H. \& {Urry}, C.~M. 2002, \apj, 579, 530

\end{thebibliography}

\end{document}